\documentclass[superscriptaddress,superscriptbib]{revtex4-2}
\usepackage{dirtytalk}
\usepackage{graphicx}  
\usepackage{amsmath}
\usepackage{mathtools}
\usepackage{amsthm}
\usepackage{amssymb}
\usepackage{bbold} 
\usepackage{mathrsfs}
\usepackage{array}
\usepackage[dvipsnames,usenames]{color}
\usepackage{soul} 
\usepackage{ulem} 
\usepackage{physics}
\usepackage[
    colorlinks=true,
    linkcolor=blue,
    citecolor=blue,
    urlcolor=blue
]{hyperref}
\usepackage{titlesec}

\titleformat{\section}
  {\normalfont\large\bfseries\raggedright}
  {}
  {0pt}
  {}

\titleformat{\subsection}
  {\normalfont\normalsize\bfseries\raggedright}
  {}
  {0pt}
  {}

\titleformat{\section}
  {\normalfont\large\bfseries\raggedright}
  {}
  {0pt}
  {}

\titleformat{\subsection}
  {\normalfont\normalsize\bfseries\raggedright}
  {}
  {0pt}
  {}

\renewcommand{\figurename}{\textbf{Fig.}}
\makeatletter
\renewcommand{\fnum@figure}{\textbf{\figurename~\thefigure}}
\makeatother

\renewcommand{\tablename}{\textbf{Table}}
\makeatletter
\renewcommand{\fnum@table}{\textbf{\tablename~\thetable}}
\makeatother

\begin{document} 

\normalem

\title{Multiple Topological Haldane Phases for Symmetry-Protected Quantum Information Processing 
}

\author{João Pedro Gama D'Elia$^{*}$}
\affiliation{School of Physics, Engineering and Technology, University of York, York, United Kingdom}
\affiliation{Instituto de Física, Universidade Federal do Rio de Janeiro, Rio de Janeiro, RJ, 21941-972, Brazil}

\author{Irene D'Amico} 
\affiliation{School of Physics, Engineering and Technology, University of York, York, United Kingdom}

\author{Thereza Paiva} 
\affiliation{Instituto de Física, Universidade Federal do Rio de Janeiro, Rio de Janeiro, RJ, 21941-972, Brazil}


\begin{abstract}
Symmetry-protected topological phases have attracted significant interest at the fundamental level and as a potential platform for quantum information processing, owing to their protected edge states and resilience to perturbations. Applying these features for practical and efficient quantum computation is highly desirable, but remains an open challenge. Here, we demonstrate the partitioning into multiple independent Haldane phase subsystems of a single spin-1/2 ladder system and propose this as a scalable architecture for gate-based quantum computation, which takes advantage of the symmetry-protected topological order.  We encode qubits in the two topological states of the 
$S^{z}=0$ sector of each subsystem. Finite-size effects, typically viewed as detrimental, instead provide a controllable energy splitting that enables single-qubit rotations using only local magnetic fields. An Ising-type interaction between neighboring subsystem edges generates entangling gates, enabling universal quantum computation driven by two control parameters that are easily accessible experimentally. Our results demonstrate how symmetry-protected topological phases can be directly harnessed for circuit-model quantum computation in realistic systems.
\end{abstract}

\maketitle
\noindent
*Corresponding author: \href{mailto:joaopedro@pos.if.ufrj.br}{joaopedro@pos.if.ufrj.br}\\
\noindent
Author emails: \href{mailto:joaopedro@pos.if.ufrj.br}{joaopedro@pos.if.ufrj.br}; \href{mailto:irene.damico@york.ac.uk}{irene.damico@york.ac.uk}; \href{mailto:tclp@if.ufrj.br}{tclp@if.ufrj.br}
\newpage

\section*{Introduction}

\noindent
The ability to build and manipulate physical qubits in a highly controllable manner has advanced rapidly with the development of increasingly sophisticated quantum platforms \cite{daley2022practical, cornish2024quantum, wu2021concise, foss2025progress}. These advances enable the realization of more robust qubits, often described by Hamiltonians that go beyond simple two-level systems. Such increased complexity is typically exploited to enhance coherence times and robustness against disorder, both of which are essential requirements for scalable quantum computation and long-lived quantum memories \cite{PhysRevA.97.033823, wang2017single, 4xcd-wyxx}.
\\
\noindent
These goals naturally align with symmetry-protected topological (SPT) phases of matter. In one-dimensional systems, SPT phases are characterized by the presence of zero-energy edge states in open chains, which are protected by specific symmetries of the Hamiltonian \cite{PhysRevB.96.165124}. A paradigmatic example is the Haldane phase \cite{PhysRevLett.50.1153, HALDANE1983464}, a one-dimensional SPT phase characterized by symmetry-protected spin-1/2 edge states and a finite bulk excitation gap (the Haldane gap) \cite{tasaki2020physics, affleck2004rigorous}.
\\
The Haldane phase has been realized across several quantum simulation platforms, including ultracold atoms in ladder geometries \cite{sompet2022realizing}, trapped-ion qutrit simulators with tunable long-range interactions \cite{PhysRevX.5.021026}, and more recently qutrit systems where edge operators enable controlled rotations within the $S_z = 0$ topological sector \cite{PRXQuantum.6.020349}, exhibiting coherence and entanglement that persist for several milliseconds. 
\\
These and related works have explored the potential of SPT phases for quantum memories \cite{PhysRevResearch.2.013120} and measurement-based quantum computation (MBQC) \cite{PhysRevLett.101.010502, PhysRevLett.105.110502,PhysRevA.76.052315, PhysRevA.96.012302}. However, it was shown that one-dimensional SPT phases cannot be universal resources for MBQC by themselves \cite{else2012symmetry}.
Moreover, proposals for MBQC in 1D Haldane phases require sufficiently large system sizes so that the edge–edge coupling becomes exponentially suppressed \cite{PhysRevLett.101.010502}. In realistic finite systems, however, the residual interaction between edges lifts the degeneracy of the topological manifold, seemingly undermining its usefulness for quantum information processing. 
\\
Still, it remains an open question whether and how such systems can be harnessed for quantum computation by using gate-based (circuit-model) quantum computation. In particular, a clear identification of the physical qubits and a systematic framework for implementing universal quantum gates are still lacking.
\\
Motivated by these challenges and by the increasing experimental accessibility of SPT phases, we first demonstrate that a single spin-1/2 ladder can be converted into multiple independent Haldane-phase subsystems by controlling only local couplings between edge rungs. This model could be implemented in a host of experimentally-accessible physical systems\cite{cai2023coherent, zhang2025universal,doi:10.1126/science.1217692, wang2012composite, hirthe2023magnetically, li2023observation, bermudez2012quantum}. Second, we identify the two topological states in the $S_z=0$ sector as logical qubits, and show that universal quantum computation can be implemented within a single ladder architecture using just two experimentally-controllable interaction parameters. Furthermore, the system sizes which are currently experimentally accessible also correspond to SPT order that persists at higher temperatures (supplementary information of Ref. \cite{sompet2022realizing}). For these, we show that the finite-size splitting between the topological ground and excited states of open chains can be exploited as a resource for single-qubit gates. 
\\
Our results demonstrate the possibility of creating a scalable number of Haldane phases and pave the way for a practical implementation of quantum information processing exploiting the topological properties of these phases. 

\section*{Results}
\subsection*{Creation of a set of independent Haldane subsystems}
\begin{figure}
    \centering
    \includegraphics[width=0.7\linewidth]{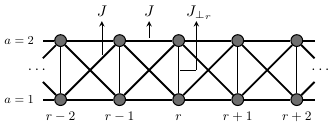}
    \caption{$|$ \textbf{Ladder system.} The two-leg Heisenberg ladder with diagonal exchange interaction. While the diagonal and leg exchange interactions are constant and equal, the rung exchange interaction \(J_{\perp_{r}}\) depends on the rung index \(r\).} 
    \label{fig:ladder_system}
\end{figure}

\begin{figure}[b]
    \centering
    \includegraphics[width=0.7\linewidth]{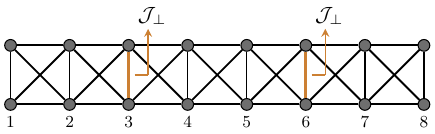}
    \caption{$|$ \textbf{Haldane subsystems via a local rung singlet.} Illustration of a ladder system with $L=8$ and $N = 2$. The rung exchange term in rungs $3$ and $6$ are given by $\mathcal{J}_{\perp}$. The system gets partitioned in $N + 1 = 3$ subsystems, separated by the $N$ rungs. As $\mathcal{J}_{\perp}$ is enhanced, the system crosses a 1QPT, where each subsystem becomes effectively independent spin-1 chains separated by the rung-singlets at positions $3$ and $6$.}
    \label{fig:ladder_several}
\end{figure}

\begin{figure}
    \centering
    \includegraphics[width=0.7\linewidth]{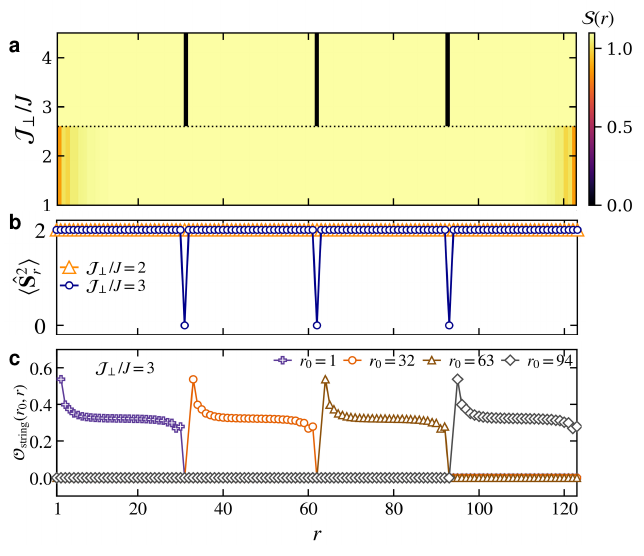}
    \caption{$|$ \textbf{First-order transition to Haldane subsystems.} \textbf{a} Bipartite entanglement entropy $\mathcal{S}(r)$ for $L=123$ and as the rung couplings $\mathcal{J}_{\perp}$ of rungs $31$, $62$, and $93$ are increased (calculations done using DMRG). The entropy between rungs $31$, $62$, and $93$ and the rest of the system vanishes beyond a critical value of $\mathcal{J}_{\perp}$, signaling that the ladder splits into four independent subsystems. \textbf{b} Expectation value $\langle \hat{\mathbf S}_r^2\rangle$ showing the transition from a uniform spin-1 chain ($\mathcal{J}_{\perp}/J=2$) to four spin-1 chains separated by spin-0 rungs ($\mathcal{J}_{\perp}/J=3$). \textbf{c} The string-order correlation function, indicator of hidden antiferromagnetic order, as a function of the rung index. It is possible to see that, for $\mathcal{J}_{\perp}>\mathcal{J}_{\perp_c}$,  the correlator is non-zero only within each subsystem.}
    \label{fig:panel1_several}
\end{figure}

The Hamiltonian of the 2-leg spin-1/2 ladder system is given by 
\begin{equation}
\begin{gathered}
\hat{H} = J\sum_{a = 1}^{2}\sum_{r=1}^{L-1}\hat{\mathbf{s}}_{a,r}\cdot \hat{\mathbf{s}}_{a,r+1} + \sum_{r = 1}^{L} J_{\perp_{r}} \hat{\mathbf{s}}_{1,r}\cdot \hat{\mathbf{s}}_{2,r}
+ J\sum_{r=1}^{L-1}\left(\hat{\mathbf{s}}_{1,r}\cdot \hat{\mathbf{s}}_{2,r+1} + \hat{\mathbf{s}}_{1,r+1}\cdot \hat{\mathbf{s}}_{2,r} \right)\,,
\label{eqn:hamiltonian}
\end{gathered}
\end{equation}
where $\hat{\mathbf{s}}_{a,r}$ is the spin-1/2 operator on leg $a$ and rung $r$ (Fig. \ref{fig:ladder_system}). The equality between leg and diagonal exchanges induces the conservation of the total rung spin \cite{honecker2000magnetization} $\hat{\mathbf{S}}_r = \hat{\mathbf{s}}_{1,r} + \hat{\mathbf{s}}_{2,r}$, allowing the above Hamiltonian (equation~(\ref{eqn:hamiltonian})) to be rewritten as \cite{honecker2016thermodynamic}
\begin{equation}
\hat{H} = J\sum_{r=1}^{L-1} \hat{\mathbf{S}}_r\cdot\hat{\mathbf{S}}_{r+1} +  \sum_{r=1}^{L} J_{\perp_{r}}\left(\frac{\hat{\mathbf{S}}_{r}^{2}}{2} - \frac{3}{4} \right) \,.
\label{eqn:T_hamiltonian}
\end{equation}
 Hence, the rules of angular momentum addition give $\langle\hat{\mathbf{S}}_{r}^{2}\rangle = S_r(S_r +1)$ with $S_r\in\{0,1\}$. This mapping is the key ingredient of our architecture: by locally tuning $J_{\perp_{r}}$, we can choose whether a given rung behaves as an effective spin-1 or spin-0 site. 
Indeed, starting with a ladder with $L$ rungs and uniform rung couplings ($J_{\perp_{r}}=J=1$ for all $r$), the ground state of the system is composed of rung-triplets, such that the ground state of the Hamiltonian (equation~(\ref{eqn:T_hamiltonian})) is the ground state of a spin-1 Heisenberg chain.
\\
Let us now select $N$ non-consecutive rungs located away from the system edges ($r\ne1,\,L$). By locally enhancing the rung coupling on these sites, indicated by $\mathcal{J}_{\perp}$ in Fig.~\ref{fig:ladder_several}, the system undergoes a first-order quantum phase transition (1QPT):  the selected rungs are driven to a rung-singlet state ($S=0$), decoupling neighboring regions and effectively partitioning the ladder into $N+1$ independent spin-1 Heisenberg chains.
In fact, if the chosen $N$ rungs partition the system into subsystems of equal size, we can show that the critical value for the rung coupling $\mathcal{J}_{\perp}$ is
\begin{equation}
    \mathcal{J}_{\perp_{c}}(L, N) = \frac{1}{N}\left[\left(N +1\right) E_\text{H}\left( T \right) - E_\text{H}(L) \right]\,,
    \label{eqn:jcrit_equal}
\end{equation}
where $E_\text{H}(l)$ is the ground-state energy of the spin-1 Heisenberg chain with $l$ sites, and $T = (L - N)/(1 + N)$ is the size of the subsystems partitioned by the $N$ rungs (see Supplementary Note 1). 
\\
This transition can be directly observed by looking at the entanglement structure. The bipartite Von Neumann entropy between each rung and the rest of the system is given by
\begin{equation}
    \mathcal{S}(r) = -\text{Tr}(\rho_r \ln{\rho_r})\,,
    \label{eqn:VN}
\end{equation}
where $\rho_r$ is the reduced density matrix of the rung $r$. As $\mathcal{J}_{\perp}$ crosses the critical value (given by equation~(\ref{eqn:jcrit_equal})) the entanglement entropy of the $N$ selected rungs drops to zero (Fig.~\ref{fig:panel1_several}a), showing that each rung-singlet acts as a boundary. Consistently, $\langle \hat{\mathbf{S}}^2_r\rangle$ changes from $2$ on all rungs (effective spin-1 chain) to a configuration with $N$ $S=0$ defects separating two spin-1 segments (Fig.~\ref{fig:panel1_several}b). 
\\
The subsystems separated by the rung-singlets become independent spin-1 Haldane phase chains. This can be seen from the string-order correlation function \cite{rspn-4cyr, PhysRevB.78.224402} given by
\begin{equation}
    \label{eqn:string_order}
    \mathcal{O}_\text{string}(r_0,r) = - \left\langle \hat{S}^z_{r_0} \exp\left(\sum_{l = r_0+1}^{r-1} i\pi \hat{S}^z_{l} \right)\hat{S}^z_{r} \right\rangle \,,
\end{equation}
and indicating the presence of hidden antiferromagnetic order \cite{PhysRevB.45.304} within each subsystem (Fig.~\ref{fig:panel1_several}c).
Thus, in the protocol proposed here, spatial modulation of the rung exchange term provides a mechanism for defining multiple
independent Haldane phase subsystems within a ladder.
\\
This architectural ingredient plays a central role in the remainder of the paper. In the following sections, we show (i) how each effective Haldane chain acts as a physical qubit, (ii) how finite-size effects enable single-qubit control, and (iii) how interactions between neighboring subsystems generate entangling gates.

\subsection*{Low-energy projection}

\begin{figure}
    \centering
    \includegraphics[width=0.7\linewidth]{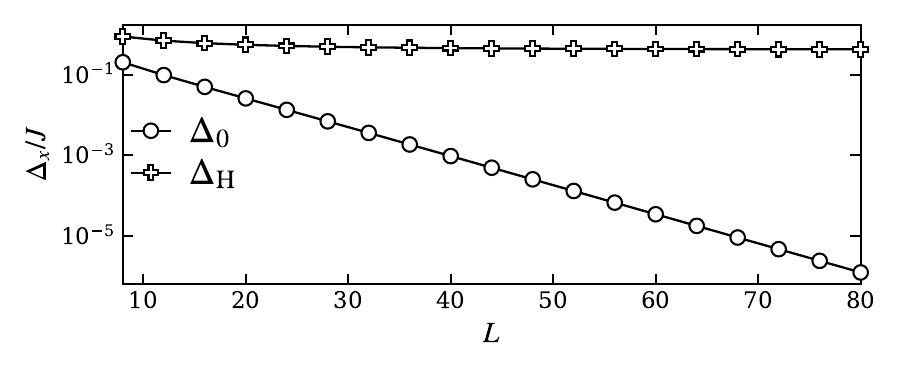}
    \caption{$|$ \textbf{Energy gaps of the topological subspace.} Size dependence of the Haldane gap $\Delta_{\mathrm H}$ and of the singlet–triplet splitting $\Delta_0$ obtained from DMRG. While $\Delta_0$ decreases exponentially with system size due to edge decoupling, the bulk Haldane gap remains nearly constant. Here $L$ refers to the size of one of the Haldane subsystems.}
    \label{fig:gap_plot}
\end{figure}

Throughout this work, we will consider each effective spin-1 Haldane subsystem as our physical qubit, described by the following Hamiltonian: \begin{equation}
\hat{H}_0 = J\sum_{r} \hat{\mathbf{S}}_{r}\cdot \hat{\mathbf{S}}_{r+1}\,. 
\end{equation}
The low-energy topologically non-trivial subspace of $\hat{H}_0$ is given by four states $\{|0\rangle,|1\rangle, |2\rangle, |3\rangle\}$ which differ only by their effective spin-1/2 edge properties \cite{Tasaki2020}. In particular, the ground-state $
|0\rangle$ is a singlet state, and the remaining three states are degenerate and correspond to triplet states, with $S_z=0$,  $+1$, and $-1$ for states $|1\rangle$, $|2\rangle$, and $|3\rangle$ respectively. These are separated from the singlet state by an energy gap $\Delta_0$ that is size-dependent and goes to zero in the thermodynamic limit, where the edges become decoupled \cite{Tasaki2020, YAMAMOTO1997545, PhysRevB.52.12844}.
\\
These four states possess two fractionalized spins-1/2 edges that are exponentially localized and weakly coupled through the gapped bulk \cite{Tasaki2020, YAMAMOTO1997545, PhysRevB.52.12844}.  Importantly, while the exponential splitting $\Delta_0$ vanishes in the thermodynamic limit, in finite systems it provides a natural energy scale that can be exploited for coherent control using it as a drift Hamiltonian enabling qubit rotations (Fig.~\ref{fig:gap_plot}, circle markers). This exponential decrease in $\Delta_0$ is in stark contrast with the Haldane gap $\Delta_\text{H}$, which has little dependence on system size, providing topological protection even for systems far from the thermodynamic limit (Fig.~\ref{fig:gap_plot}, plus markers).
In this low-energy manifold, these states can be represented by their effective spin-1/2 edge states on the left edge (L) and right edge (R) as
\begin{equation}
    \begin{gathered}
        |0\rangle = \frac{1}{\sqrt{2}}\left(|\uparrow\rangle_{L}|\downarrow\rangle_{R} - |\downarrow\rangle_{L}|\uparrow\rangle_{R} \right) \\
        |1\rangle = \frac{1}{\sqrt{2}}\left(|\uparrow\rangle_{L}|\downarrow\rangle_{R} + |\downarrow\rangle_{L}|\uparrow\rangle_{R} \right)\\
        |2\rangle = |\uparrow\rangle_{L}|\uparrow\rangle_{R}\,,\quad |3\rangle = |\downarrow\rangle_{L}|\downarrow\rangle_{R}
    \label{eqn:low_energy}
    \end{gathered}
\end{equation}
where $|\alpha\rangle_{L(R)}$, with $\alpha \in \{\uparrow, \downarrow\}$, is associated with the state of the exponentially localized spin-1/2 of the left(right) edge.
\\
These emergent edge degrees of freedom constitute the resource we propose for encoding quantum information.

\subsection*{Numerical methods}

We employ the density matrix renormalization (DMRG) to calculate the ground state and the triplet of excited states of the ladder Hamiltonian, and use time evolving block decimation (TEBD) to evaluate the time evolution of these states under different control parameters and to compare the result with the analytical results from the low-energy manifold projection. In Figures~\ref{fig:panel2}e, ~\ref{fig:panel2}f, ~\ref{fig:panel3}e, ~\ref{fig:panel3}f and ~\ref{fig:panel3}g continuous lines are the analytical results from the projection into the low-energy subspace, whereas the markers are associated with the evolution of the system with TEBD. We also use the time-dependent variational principle (TDVP) for quenches involving interactions with the spin-0 rung-singlet boundary.

\subsection*{Single-qubit control}
Having identified the topological subspace, we now show how coherent control emerges naturally from the finite-size physics of the Haldane chain. We define our physical qubit in the $P_0 = \{|0\rangle, |1\rangle\}$ subspace, with the system initialized in the ground-state $|0\rangle$, or, in general, within this subspace, and with external control terms which do not mix the $S_z = 0$ with the $S_z = \pm 1$ subspace given by $P_{\pm} = \{|2\rangle, |3\rangle\}$. This corresponds to the addition of a control interaction term to the Hamiltonian $\hat{H}_0$ block-diagonal in the low-energy subspace basis, with one block in $P_0$ and another in $P_{\pm}$. This guarantees that the system will only evolve within $P_0$.
\begin{figure*}[t]
  \centering
  \includegraphics[width=\textwidth]{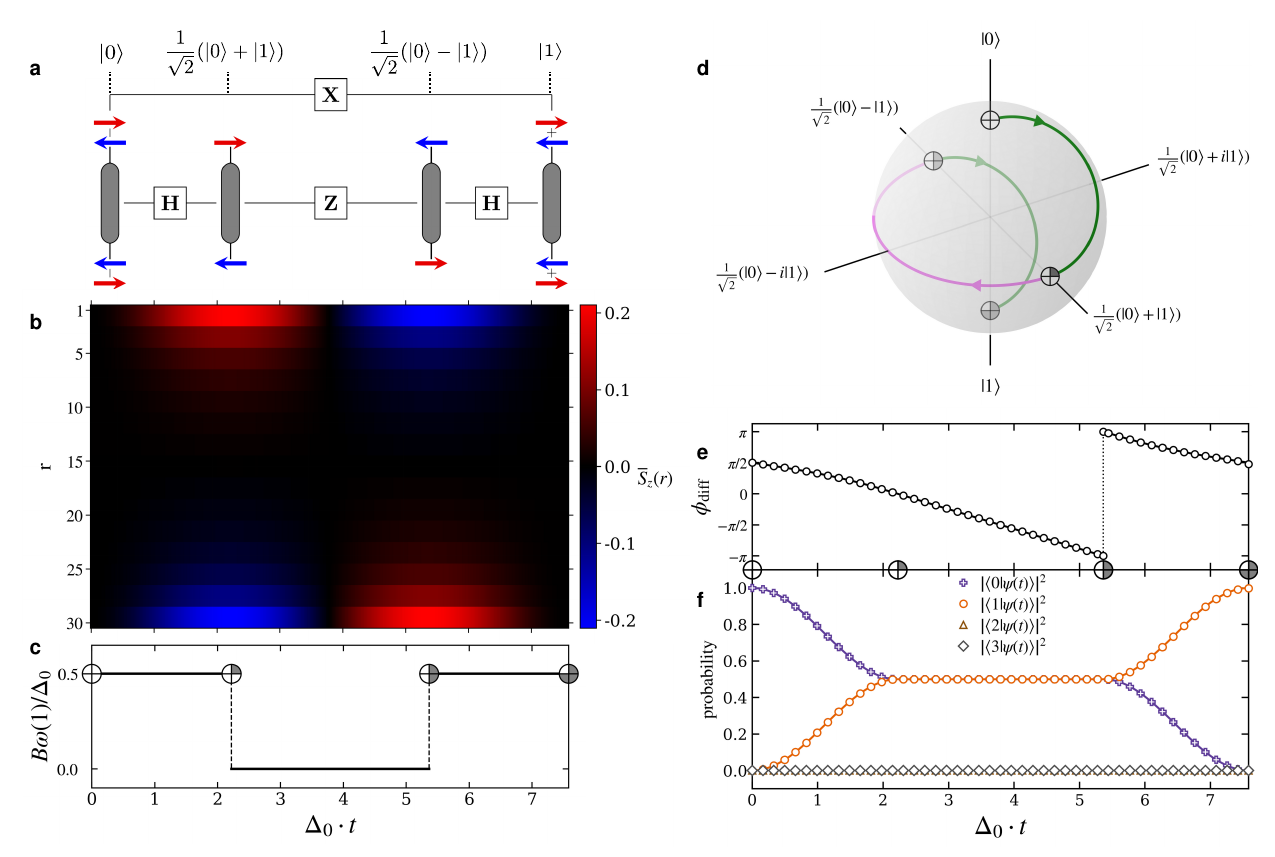}
  \caption{$|$ \textbf{Single-qubit gate implemented using finite size splitting and local control.} \textbf{a} Gate sequence realizing a Pauli-X operation via two Hadamard rotations and a Pauli-Z operation. \textbf{b} Time evolution of the edge magnetization profile during the protocol (TEBD) for a system with $L=30$ rungs. The vertical axis corresponds to the rung index, displaying the sum magnetization given by equation \ref{eqn:edge_mag}. \textbf{c} Control sequence for the boundary magnetic field. \textbf{d} Trajectory of the Bloch vector in the logical subspace. The green curves correspond to $B\ne0$ intervals and the pink curve to $B=0$. The path was built using the TEBD data of panels \textbf{e} and \textbf{f}. \textbf{e} Relative phase evolution between $|0\rangle$ and $|1\rangle$. \textbf{f} Population dynamics within the low-energy manifold. The solid lines in panels \textbf{e} and \textbf{f} are given by equation~(\ref{eqn:time_evolution_single}) while symbols correspond to the evolution of a system with $L=30$.
  }
  \label{fig:panel2}
\end{figure*}
The simplest term that obeys this condition is any term proportional to $\hat{S}^{z}_r$, where in the $P_0$ subspace we have
\begin{equation}
    \hat{P}_0 \hat{S}_r^{z} \hat{P_0} = w_r\hat{\sigma}_x, \quad w_r = \langle 1|\hat{S}_r^{z}|0\rangle \,,
\end{equation}
with $\hat{P}_0$ the projection operator in subspace $P_0$ and $\hat{\sigma}_x$  the Pauli-$x$ operator. 
\\
Consider then a system with a control parameter given by a magnetic field $B$ on a set of sites of a Haldane subsystem
\begin{equation}
\hat{H} = \hat{H}_0 - B\sum_{r=1}^{q}\hat{S}^{z}_{r}\,, \quad 1 \le q < L\,. 
\end{equation}
In the $P_0$ subspace, the total Hamiltonian is given by
\begin{equation}
    \hat{P}_0\hat{H}\hat{P}_0 = -\frac{\Delta_{0}}{2}\hat{\sigma}_z - B\omega(q)\hat{\sigma}_x = \mathbf{h}\cdot \boldsymbol{\sigma}
    \label{eqn:single_projection}
\end{equation}
where $\omega(q) = \sum_{r=1}^{q} w_{r}$, $\Delta_0$ is the gap between $|0\rangle$ and $|1\rangle$, $\mathbf{h} = (-B\omega(q), 0 , -\Delta_{0}/2)$ and $\boldsymbol{\sigma} = (\hat{\sigma}_{x}, \hat{\sigma}_{y}, \hat{\sigma}_{z})$. 
The time evolution operator of the system in the $P_0$ subspace is given by
\begin{equation}
\label{eqn:time_evolution_single}
    \hat{\mathcal{U}}_0(t) = 
    \begin{cases}
        e^{i\frac{\Delta_{0} t}{2}\hat{\sigma}_z}, \,\,\,\,\quad B = 0\\
        e^{-i |\mathbf{h}|t\, \hat{\mathbf{h}} \cdot \boldsymbol{\sigma}}, \,\, B \ne 0\,.
    \end{cases}
\end{equation}
The above operators are rotation operators. In particular, for $B = 0$, the time evolution of the system for a time $t$ is given by the rotation of the Bloch vector around the $-\hat{\mathbf{z}}$ axis by an angle of $\Delta_{0} t$, and for $B \ne 0$ the time evolution is equivalent to rotations around $\hat{\mathbf{h}}$ by an angle $2|\mathbf{h}|t$, where $\hat{\mathbf{h}}$ is a unitary vector in the $xz$ plane. Then the magnetic field $B$ (control parameter) and the gap $\Delta_0$ between $|0\rangle$ and $|1\rangle$ (drift parameter) provide two non collinear axes of rotations, which, through composition, generate the SO(3) rotation group as a direct consequence of Euler's rotation theorem \cite{gothen2023euler}: the control plus drift parameters allow the implementation of arbitrary single-qubit gates.

As an explicit demonstration, Fig.~\ref{fig:panel2} implements a Pauli-X gate using a sequence of two Hadamard rotations (Fig.~\ref{fig:panel2}a) generated by a finite field quench and a relative phase accumulation generated by free evolution at $B=0$ (Fig.~\ref{fig:panel2}c). In this setting, we demonstrate the creation of unitary gates using a magnetic field applied only to the leftmost rung of the Haldane subsystem ($q=1$ in equation~(\ref{eqn:single_projection})), as a minimal setup allowing single-qubit rotations.

The Hadamard gates can be constructed by applying $B = \Delta_{0}/2\omega(q)$ for a time interval of $t = \sqrt{2}\pi/2\Delta_{0}$ that accounts for a rotation of an angle of $\pi/2$ around the unit vector $-\frac{1}{\sqrt{2}}(\hat{x} + \hat{z})$, whereas the Pauli-Z gate comes from the natural evolution of the system ($B=0$) through a rotation of $\pi$ around $-\hat{z}$ for and interval of $t = \pi/\Delta_{0}$ (Figs. \ref{fig:panel2}d, ~\ref{fig:panel2}e and \ref{fig:panel2}f).

The gate action in the Haldane phase subsystem is directly observable in the edge magnetization pattern. Due to its staggered pattern, the magnetization displayed in Fig.~\ref{fig:panel2}b  is the sum of couples of consecutive rungs according to
\begin{equation}
    \overline{S}_{z}(r) = \langle S^{z}_{r} + S^{z}_{r+1} \rangle,\quad r\in\{1,3,\cdots,L-1\}.
    \label{eqn:edge_mag}
\end{equation}
Although the bulk remains unchanged, the state of each edge changes dynamically (Fig.~\ref{fig:panel2}b), consistent with the projected edge-spin representation in equation~(\ref{eqn:low_energy}).
\begin{equation}
    \begin{gathered}
        \frac{1}{\sqrt{2}}(|0\rangle + |1\rangle) = |\uparrow\rangle_{L}|\downarrow\rangle_{R}\,,\\
        \frac{1}{\sqrt{2}}(|0\rangle - |1\rangle) = |\downarrow\rangle_{L}|\uparrow\rangle_{R} \,,
    \end{gathered}
\end{equation}
which correspond to the states after the Hadamard and Hadamard plus Pauli-Z gates, respectively.

Importantly, our protocol requires only a local magnetic field acting on a single boundary rung as the minimal setting; however, if necessary, e.g., because of practical requirements, the magnetic field could be applied to several sites, provided it is not uniform across the entire chain (i.e., constant $B$ for all rungs). We have therefore established full single-qubit controllability using only local boundary fields. We now show that neighboring Haldane subsystems can be coupled through their edge spins to generate entangling operations.

\subsection*{Two-qubit entangling gates}

\begin{figure*}[t]
  \centering
  \includegraphics[width=\textwidth]{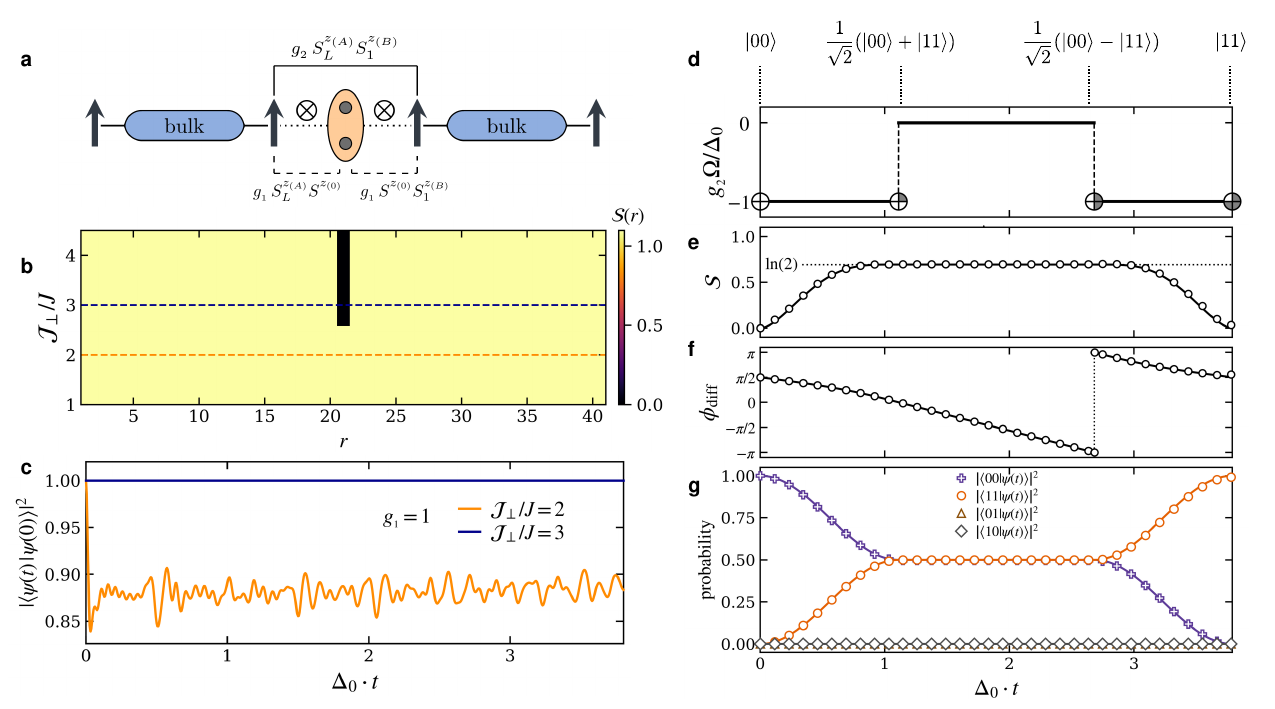}
  \caption{$|$ \textbf{Entangling gate.} \textbf{a} The edge rungs of two neighboring Haldane subsystems (A and B), with $L=20$ rungs each, coupled through an Ising-type interaction between their boundary rungs (solid line with $g_{_2}$) and between the boundary rungs and the rung-singlet (dashed lines with $g_{_1}$). \textbf{b} Bipartite entanglement entropy $\mathcal{S}(r)$ as the rung coupling $\mathcal{J}_{\perp}$ is increased (DMRG, $L=41$). The entropy across the rung $21$ vanishes beyond a critical value, splitting the ladder into two independent subsystems. The dashed lines mark the values of $\mathcal{J}_{\perp}$ used in panel c for the quenches with $g_{_1}$. \textbf{c} The fidelity with the initial state as a function of time during a quench of $g_{_1} = 1$ (TDVP). \textbf{d} Quench protocol for the edge-edge coupling $g_{_2}$. \textbf{e} Entanglement entropy between the two subsystems showing the creation and destruction of entangled states. The dotted line marks the maximum value of entanglement: $\ln{(2)}$. \textbf{f} Relative phase between states $|00\rangle$ and $|11\rangle$. \textbf{g} Population dynamics within the $P$ subspace.}
  \label{fig:panel3}
\end{figure*}

To scale beyond single-qubit control, our protocol requires an interaction between neighboring Haldane subsystems. Because logical information resides in the fractionalized edge spins, it is sufficient to couple only the boundary rungs of adjacent subsystems, leaving the gapped bulk intact, and constraining any interaction term to the logical subspace $P = P_0 \times P_0$.
One way to do this is to consider an Ising-type interaction between two consecutive edges of the physical qubits (Fig.~\ref{fig:panel3}a, solid line with $g_{_2}$), given by the Hamiltonian
\begin{equation}
    \hat{H} = g_{_2}\cdot \hat{S}^{z_{(A)}}_{1} \hat{S}^{z_{(B)}}_{L} + \hat{H}_0^{_{(A)}}\otimes\hat{\mathbb{1}} + \hat{\mathbb{1}}\otimes\hat{H}_0^{_{(B)}}\,,
    \label{eqn:two_qubit_hamiltonian}
\end{equation}
where the superscript $A$($B$) refers to a left(right) Haldane subsystem. In the projected subspace, this interaction is given by $g_{_2}\Omega\cdot\hat{\sigma}_x\otimes\hat{\sigma}_x$, where
\begin{equation}
\Omega = \langle 0_{(A)}| \hat{S}^{z_{(A)}}_{1}|1_{(A)}\rangle \langle 0_{(B)}| \hat{S}^{z_{(B)}}_{L}|1_{(B)}\rangle  \,.    
\end{equation}
Alongside the single-qubit Hamiltonians for systems A and B, which in the low-energy projection is given by the gap between states $|0\rangle$ and $|1\rangle$,  this gives 
\begin{equation}
    \hat{P}\hat{H}\hat{P} = g_{_2}\Omega\cdot\hat{\sigma}_x\otimes\hat{\sigma}_x -\frac{\Delta^{_{(A)}}_{0}}{2}\hat{\sigma}_z\otimes\hat{\mathbb{1}} -\frac{\Delta^{_{(B)}}_{0}}{2}\hat{\mathbb{1}}\otimes\hat{\sigma_z}
\end{equation}
On the basis of $P$, the projected Hamiltonian matrix becomes the block diagonal $ H = H_{\{|00\rangle,|11\rangle\}} \oplus H_{\{|01\rangle,|10\rangle\}}$, where 
\begin{equation}
\begin{aligned}
    H_{\{|00\rangle,|11\rangle\}} =
    \begin{pmatrix}
        -\frac{(\Delta^{_{(A)}}_{0} + \Delta^{_{(B)}}_{0})}{2} & g_{_2}\Omega \\
        g_{_2}\Omega & \frac{(\Delta^{_{(A)}}_{0} + \Delta^{_{(B)}}_{0})}{2}
    \end{pmatrix},\quad
    H_{\{|01\rangle,|10\rangle\}} =
    \begin{pmatrix}
        \frac{(\Delta^{_{(B)}}_{0} - \Delta^{_{(A)}}_{0})}{2} & g_{_2}\Omega \\
        g_{_2}\Omega & \frac{(\Delta^{_{(A)}}_{0} - \Delta^{_{(B)}}_{0})}{2}
    \end{pmatrix},
\end{aligned} 
\end{equation}
leading to time evolving operators with the same structure as the one defined in equation~(\ref{eqn:time_evolution_single}) but with $g_{_2}$ as the control parameter. This interaction acts entirely within the logical subspace and provides the entangling resource required for universal quantum computation.
In order to create an entangling two-qubit gate in our system, controlled by $g_{_2}$, we need to perform a next-nearest neighbor interaction between boundary rungs, as these are separated by a rung-singlet. In quantum simulation platforms such as trapped ion systems, this Ising exchange coupling can be generated with the application of microwave or optical fields \cite{RevModPhys.93.025001, foss2025progress, zygelman2021first,kim2011quantum}, with a controllability of both intensity and sign of nearest-neighbors and next-nearest-neighbors interactions \cite{kim2010quantum}. Similar capabilities exist in neutral-atom arrays, where parallel entangling operations between distant atoms can also be realized \cite{evered2023high,henriet2020quantum}.
\\
Although entangling interactions between specific ion pairs beyond nearest neighbors have already been demonstrated, suppressing or canceling residual exchange interactions with neighboring ions when implementing such gates remains challenging \cite{debnath2016demonstration, figgatt2019parallel, RevModPhys.93.025001, PhysRevA.94.042308}. 
We therefore are going to consider the possibility that the entangling gate generates an additional Ising-type interaction with the neighboring rung singlet (Fig.~\ref{fig:panel3}a, dashed lines with $g_{_1}$). The ladder architecture plays a fundamental role here: any spin interaction involving a rung singlet vanishes, since the spin-0 representation of $SU(2)$ is trivial and its spin operators are zero.

Figure~\ref{fig:panel3}b shows the von Neumann entropy of each rung (equation~(\ref{eqn:VN})) for a system of $L=41$ rungs as the middle-rung exchange increases, revealing the 1QPT. The two values of $\mathcal{J}_{\perp}/J$ highlighted by the dashed lines are associated with the quenches with $g_{1}=J=1$ shown in Fig.~\ref{fig:panel3}c. For $\mathcal{J}_{\perp}/J=2$, below the critical value,  the ladder is entirely within a single Haldane phase, and the fidelity with the initial state oscillates and decreases in time. For $\mathcal{J}_{\perp}/J=3$, above the critical point, the fidelity remains unity, demonstrating robustness against exchange interactions with the rung-singlet boundaries.

Thus, we can show that the Hamiltonian equation~(\ref{eqn:two_qubit_hamiltonian}) generates coherent oscillations within the logical two-qubit manifold, allowing controlled creation of entanglement. Figure~\ref{fig:panel3}d shows that a simple quench of the edge coupling $g_{_2}$ periodically produces and removes entanglement between the subsystems: the bipartite entanglement entropy between the two Haldane subsystems reaches its maximum value at well-defined times (Fig.~\ref{fig:panel3}e), while the corresponding phase and population dynamics (Fig.~\ref{fig:panel3}f and ~\ref{fig:panel3}g) confirm the formation of Bell states within $P=P_0\times P_0$.

To quantify the performance of these operations beyond the agreement between the low-energy projection and the full many-body time evolution, we compute the gate fidelity and leakage probability for representative single- and two-qubit gates. The gate fidelity is defined as \(F_{\mathrm{gate}}=|\langle \psi_{\mathrm{id}}|\psi(t_{\mathrm{gate}})\rangle|^2\),
where \(|\psi_{\mathrm{id}}\rangle\) is the target state obtained from the ideal logical operation. The leakage probability is \(P_{\mathrm{leak}}=1-\langle \psi(t_{\mathrm{gate}})|\hat P_{\mathrm{log}}|\psi(t_{\mathrm{gate}})\rangle\), where \(\hat P_{\mathrm{log}}\) projects onto the logical subspace. Representative values are summarized in Table I, with gate times reported in dimensionless units. A detailed discussion of the definitions, projection procedure, and parameter dependence is provided in Supplementary Note 2 and Supplementary Tables S1 and S2.
\begin{table}
\label{tab:main_operations}
\setlength{\tabcolsep}{6pt}
\begin{tabular}{cccccc}
\hline
Operation & System & Control & Gate time ($J^{-1}$) & \(F_{\mathrm{gate}}\) & \(P_{\mathrm{leak}}\) \\
\hline
Single-qubit \(H\) gate & \(L=30\) & boundary field \(B/J = 4.70\cdot10^{-3}\) & $443.4$ & $0.99998$ & $1.56 \cdot 10^{-5}$ \\

Bell-state generation & \(L_A=L_B=30\) & edge coupling \(g_2/J = 1.76 \cdot 10^{-2}\) & $221.7$ & $0.99986$ & $1.38 \cdot 10^{-4}$ \\
\hline
\end{tabular}
\caption{$|$ \textbf{Representative gate performance.}
Gate fidelities and leakage probabilities obtained from the full many-body time
evolution for the single-qubit and two-qubit protocols of the biggest system sizes simulated. Here \(F_{\mathrm{gate}}\) measures the overlap with the ideal logical target
state, while \(P_{\mathrm{leak}}\) measures the probability of leaving the
logical subspace. Times are given in dimensionless units, with \(\hbar=1\) and
\(J=1\).}
\end{table}

These results show that the proposed control protocols implement the desired
logical operations while keeping the many-body state confined to the
topological qubit manifold. The combination of arbitrary single-qubit rotations
and an entangling two-qubit operation therefore provides the ingredients for
universal gate-based quantum computation within the partitioned Haldane ladder \cite{PhysRevLett.89.247902, PhysRevA.65.040301}.

\section*{Discussion}

The possibility of building several symmetry-protected qubits, including of physically different sizes, in a single ladder structure opens new paths in designing new quantum computation and quantum memory platforms possessing both enhanced robustness and boosted coherence times. We have introduced a scalable protocol for universal quantum computation based on the Haldane phase on a spin-1/2 ladder. By spatially modulating the rung coupling, the ladder can be partitioned into multiple independent Haldane subsystems within a single device, each providing a protected topological qubit encoded in the $S_z=0$ sector of the low-energy manifold.

A central feature of our approach is the deliberate use of finite-size effects. The residual coupling between edge spins, often regarded as a limitation, generates an intrinsic energy splitting that acts as a drift Hamiltonian enabling coherent control. When combined with a local magnetic field acting on a boundary rung, this splitting provides two non-collinear rotation axes, allowing arbitrary single-qubit gates using minimal local control. Entangling operations arise from an Ising-type interaction between neighboring subsystem edge rungs, demonstrating the ability to create maximally entangled states using short-range interactions, independently of the physical qubits' size. Together, these ingredients satisfy the requirements for universal gate-based quantum computation.

Our results show that symmetry-protected topological phases can be directly harnessed for circuit-model quantum computing in systems and setups available within current technology. Beyond quantum computation, the proposed architecture highlights a broader paradigm in which finite-size effects in topological phases become a resource rather than a limitation. Moreover, the ladder Hamiltonian, together with its local symmetries of rung spin conservation, allows the creation of Haldane subsystems of varied size within the same ladder, allowing the construction of different architectures for encoding quantum information, which will be explored in future works. We anticipate that this approach will stimulate further exploration of SPT phases as versatile platforms for quantum technologies.

\section*{Methods}

We studied the system using different matrix-product-state methods with calculations done using the TeNPy library \cite{tenpy2024}. All the eigenstates in the low-energy topological manifold were obtained with finite-system density-matrix renormalization group calculations using open boundary conditions. Unless otherwise stated, the energy scale was set by J=1. For the ground state $\ket{0}$ and excited states $\ket{2}$ and $\ket{3}$ calculations were performed in the $S^{z}_\text{tot} = 0\,, +1\,\text{and}\, -1$ sectors, respectively. The $\ket{1}$ state was obtained by performing an orthogonal excitation on state $\ket{0}$. The maximum MPS bond dimension for all state calculations was increased up to $\chi_\text{max} = 600$ with at most $30$ sweeps and a truncation error of $10^{-8}$.
\\
Real-time dynamics were simulated using both time-evolving block decimation and time-dependent variational principle methods. For TEBD, we used a fourth-order Suzuki–Trotter decomposition with time step $dt=0.05$ and maximum bond dimension $\chi_\text{max} = 600$. For TDVP, time evolution was performed using the single-site algorithm with time step $dt=0.1$. The bond dimension was allowed to grow up to $\chi_\text{max} = 600$.

\section*{Data Availability}
The data underlying the figures in this article have been deposited in Zenodo and will be made publicly available upon publication of the article. For peer review, access is provided through a private Zenodo review link. The final dataset will be available at DOI: 10.5281/zenodo.20627836.

\bibliographystyle{apsrev4-2}
\bibliography{manuscript}

\section*{Acknowledgements}
T.P.~acknowledges support from Funda\c c\~ao Carlos Chagas de Apoio \`a Pesquisa (FAPERJ) Grants No.~E-26/210.100/2023, E-26/210.781/2025, and   E-26/200.230/2026, and from Conselho Nacional de Desenvolvimento Cient\'\i fico e Tecnol\'ogico (CNPq), grant numbers 308335/2019-8 and 442072/2023-6. J.P.G.D. acknowledges support from CNPq grant number 442072/2023-6 and from Coordenação de Aperfeiçoamento de Pessoal de Nível Superior (CAPES). J.P.G.D. and T.P. thank the School of Physics, Engineering and Technology, University of York (United Kingdom) for the kind hospitality.

\section*{Author Contributions}
J.P.G.D. and I.D. conceived the project. J.P.G.D. performed the analytical and numerical calculations.  J.P.G.D., I.D., and T. P. analysed the data, discussed the contents, and wrote the manuscript. 

\section*{Competing Interests}
The authors declare no competing interests.

\newpage

\onecolumngrid
\setcounter{equation}{0}
\setcounter{figure}{0}
\setcounter{table}{0}
\renewcommand{\theequation}{S\arabic{equation}}
\renewcommand{\thefigure}{S\arabic{figure}}
\renewcommand{\thetable}{S\arabic{table}}

\section*{Supplementary Information}

\section*{Supplementary Note 1: First-order quantum phase transition}

In the main text, we show that spatial modulation of the rung-exchange coupling $J_{\perp_{r}}$ allows a single ladder to host multiple independent spin-1 Haldane segments separated by spin-0 rung-singlet barriers, corresponding to the Hamiltonian
\begin{equation}
\hat{H} = J\sum_{r=1}^{L-1} \hat{\mathbf{S}}_r\cdot\hat{\mathbf{S}}_{r+1} + \sum_{r=1}^{L} J_{\perp_{r}}\left(\frac{\hat{\mathbf{S}}_{r}^{2}}{2} - \frac{3}{4} \right) \,.
\label{eqn:Tsup_hamiltonian}
\end{equation}
Here we derive the critical value of $J_{\perp_{r}}$ at which the selected rungs change from a triplet configuration $S_r = 1$ to a singlet configuration, $S_r = 0$. This transition follows from the local conservation of the total rung spin $\hat{\mathbf{S}}_{r} = \hat{\mathbf{s}}_{1,r} + \hat{\mathbf{s}}_{2,r}$, which restricts each rung to well-defined spin sectors.
\\
To demonstrate this and determine the critical coupling for a finite ladder, we begin with a few definitions, taking $J=1$ as the energy scale. Let $E_\text{H}(L)$ be the ground state energy of the spin-1 Heisenberg chain with $L$ sites (Supplementary Fig. \ref{fig:energy_jcrit}a), and $\epsilon_\text{H} = \lim_{L\rightarrow\infty}E_\text{H}(L)/L$ be the energy per site in the thermodynamic limit. If, for all $r$, $J_{\perp_{r}} < E_\text{H}(L)/L$ each rung favors being in a triplet configuration, where each rung gets mapped into a spin-1 site ($S_r = 1$), and if, for all $r$, $J_{\perp_{r}} > E_\text{H}(L)/L$ each rung favors being in a singlet configuration, getting mapped to a spin-0 site ($S_r = 0$).
\\
Now suppose that $J_{\perp_{r}} < E_\text{H}(L)/L$ on all rungs except for a subset $R$, where the rung exchange is set to a tunable value $\mathcal{J}_\perp$. We impose that the rung indices of $R$ are not consecutive and are not on the edges of the system ($\{1, L\} \notin R$). This way our ladder is a collection of $|R| + 1$ cells of rungs with exchange $J_{\perp_{_r}}< E_\text{H}(L)/L$, with $r \notin R$, separated by single rungs with exchange $\mathcal{J}_\perp$. Let $T$ be the set of integers that give the number of rungs in each one of the cells: $T = \{L_1, \cdots, L_{|R|+1}\}$ (see Supplementary Fig.~\ref{fig:ladder_sup} as an example).
\begin{figure*}[b]
    \centering
  \includegraphics[width=\textwidth]{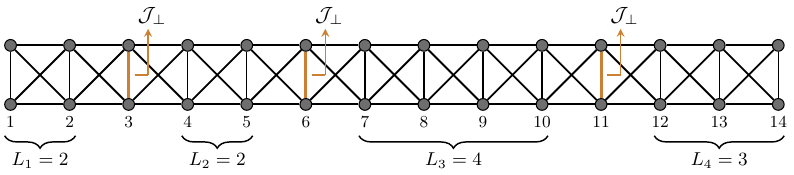}
  \caption{\textbf{Schematics of the system}. Example of a ladder system with $L = 14$ rungs (with its respective indices bellow the rungs) where the set of rungs with exchange interaction $\mathcal{J}_{\perp}$ is $R = \{3, 6, 11\}$ and the set of integers that give the number of rungs in each cell is $T = \{L_1, L_2, L_3, L_4\}$.}
  \label{fig:ladder_sup}
\end{figure*}
We compare two competing sectors: one in which all rungs remain in triplet configurations, forming a single spin-1 chain of length $L$, and another in which the rungs become singlets, thereby cutting the system into independent spin-1 chains of lengths $L_i$. 
\\
The ground state energy in the sector where all rungs have $S_r = 1$ is given by
\begin{equation}
    E_1(L, R) = E_\text{H}(L) + \sum_{r \notin R}\frac{J_{\perp_{r}}}{4} +\frac{\mathcal{J_\perp}}{4}|R|
\end{equation}
and the ground state energy in the sector where $S_r=1$ if $r\notin R$ and $S_r = 0$ if $r \in R$ is given by
\begin{equation}
    E_0(L,R,T) = \sum_{L_i \in T}E_\text{H}(L_i) + \sum_{r\notin R}\frac{J_{\perp_{r}}}{4} - \frac{3\mathcal{J_{\perp}}}{4}|R|
\end{equation}
Then the gap between these sectors is given by $E_1(L, R) - E_0(L,R,T)$ 
\begin{equation}
    \Delta(L,R,T) = E_\text{H}(L) - \sum_{L_i \in T}E_\text{H}(L_i) + |R|\mathcal{J}_{\perp}
\end{equation}
The transition between both sectors occurs when the gap goes to zero, where $\mathcal{J}_{\perp_{c}} = \mathcal{J}_{\perp}|_{\Delta=0}$. Then 
\begin{equation}
    \mathcal{J}_{\perp_{c}}(L, R, T) = \frac{1}{|R|}\left( \sum_{L_i \in T}E_\text{H}(L_i) - E_\text{H}(L) \right)\,.
\end{equation}
For cells of the same size $N$, which are the ones we use in our work, we have the following.
\begin{equation}
    N = \frac{L-|R|}{|T|}
\end{equation}
Then the critical value is given by
\begin{equation}
    J_{\perp_{c}}(L, R, T) = \frac{1}{|R|}\left( |T|E_\text{H}(N) - E_\text{H}(L) \right)\,.
    \label{eqn:jcrit}
\end{equation}
\begin{figure}
    \centering
    \includegraphics[width=0.7\linewidth]{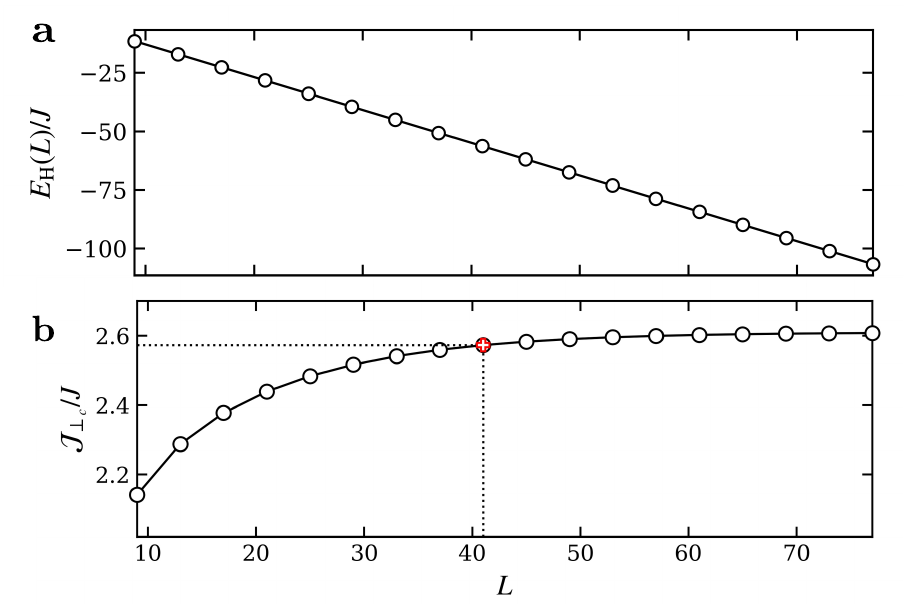}
    \caption{Panel {\bf a}: Ground state energy of the spin-1 Heisenberg chain  as a function of $L$. Panel {\bf b}: Critical value for the middle
rung exchange for a system with two cells of the same
size, as a function of $L$. The red plus marker is the critical value of the 1QPT found directly through DMRG for the system with $L=41$ rungs studied in Figure 6 of the main paper.  }
    \label{fig:energy_jcrit}
\end{figure}
In Fig. \ref{fig:energy_jcrit}b, we show the critical value for the middle rung exchange for a system with two cells of the same size, calculated using equation \ref{eqn:jcrit}. The red cross marker is the critical value calculated through DMRG, shown in Fig. 6b of the main paper.

In the limit of large cells (spin-1 subsystems), the open spin-1 Heisenberg chain energy may be approximated as $E_\text{H}(L) = \epsilon_{H}(L-1)$, where the $-1$ accounts for the boundary correction of an open chain. Substituting this into equation~(\ref{eqn:jcrit}) we get
\begin{equation}
    J_{\perp_{c}} = -2\epsilon_\text{H}
\end{equation}
Using $\epsilon_\text{H} \approx -1.401484$, we obtain $J_{\perp_{c}} \approx 2.802968$. 

\section*{Supplementary Note 2: Gate fidelity and leakage}

In the main text, we demonstrated single-qubit rotations using the finite-size splitting of the Haldane edge states and a local boundary magnetic field, and two-qubit entangling dynamics using an Ising-type interaction between neighboring
subsystem edges. Here we provide the quantitative procedure used to evaluate the gate fidelity, leakage probability, and robustness of these protocols.
\subsection*{Logical subspace and projectors}

For a single Haldane subsystem, the logical qubit is encoded in the \(S^z=0\) sector of the low-energy topological manifold, $\mathcal{H}_{\mathrm{log}}=\mathrm{span}\{|0\rangle,|1\rangle\}$, where \(|0\rangle\) and \(|1\rangle\) are the singlet and \(S^z=0\) triplet
edge states, respectively. The corresponding logical projector is $\hat P_{0} = |0\rangle\langle 0|+|1\rangle\langle 1|$.
\\
For two neighboring Haldane subsystems, labelled \(A\) and \(B\), the logical
subspace is $\mathcal{H}_{\mathrm{log}}^{(2)} = \mathcal{H}_{\mathrm{log}}^{(A)}\otimes \mathcal{H}_{\mathrm{log}}^{(B)}$, with projector
\[
\hat P
=
\sum_{\alpha,\beta=0,1}
|\alpha_A\beta_B\rangle\langle \alpha_A\beta_B|.
\]
All overlaps and populations reported below are computed by projecting the many-body state obtained from TEBD time evolution onto these DMRG low-energy states.

\subsection*{Gate fidelity and leakage probability}

For a target logical operation \(\hat U_{\mathrm{id}}\), the ideal final state is $|\psi_{\mathrm{id}}\rangle = \hat U_{\mathrm{id}}|\psi(0)\rangle$, where $|\psi(0)\rangle$ is the initial state. The state obtained from the full many-body time evolution is $|\psi(t)\rangle = e^{-i\hat H t}|\psi(0)\rangle$. We define the gate fidelity at the gate time \(t_{\mathrm{gate}}\) as
\[
F_{\mathrm{gate}}
=
|\langle \psi_{\mathrm{id}}|\psi(t_{\mathrm{gate}})\rangle|^2 .
\]
This quantity measures the agreement between the desired logical operation and the full many-body dynamics. The leakage probability is defined as
\[
P_{\mathrm{leak}}
=
1-
\langle \psi(t_{\mathrm{gate}})|
\hat P_{\mathrm{log}}
|\psi(t_{\mathrm{gate}})\rangle ,
\]
where \(\hat P_{\mathrm{log}}\) is either \(\hat P_0\) or \(\hat P\), depending on whether a single-qubit or two-qubit protocol is considered. Thus, \(P_{\mathrm{leak}}\) measures the probability that the many-body state leaves the logical qubit subspace during the gate.
\\
The gate fidelity and the leakage probability quantify different sources of
error. To see this, the evolved state can be decomposed as
\[
|\psi(t_\text{gate})\rangle
=
\sqrt{1-P_{\mathrm{leak}}}\,
|\psi_{\mathrm{log}}(t_\text{gate})\rangle
+
\sqrt{P_{\mathrm{leak}}}\,
|\psi_{\perp}(t_\text{gate})\rangle ,
\]
where \(|\psi_{\mathrm{log}}(t_\text{gate})\rangle\) lies inside the logical subspace and \(|\psi_{\perp}(t_\text{gate})\rangle\) is orthogonal to it. Since the ideal target statebelongs to the logical subspace,
\[
F_{\mathrm{gate}}
=
(1-P_{\mathrm{leak}})
F_{\mathrm{cond}},
\]
where $F_{\mathrm{cond}} = |\langle\psi_{\mathrm{id}}|\psi_{\mathrm{log}}(t_{\mathrm{gate}})\rangle|^2$ is the conditional fidelity within the logical subspace. Therefore, leakage captures population lost from the logical manifold, whereas \(F_{\mathrm{cond}}\) captures coherent logical errors such as imperfect rotation angles or phase accumulation.

\subsection*{Single-qubit gate performance}

For the single-qubit protocol of Fig.~5, the target operation is a Pauli-X gate
constructed as a sequence of two Hadamard rotations and a Pauli-Z rotation. The
Hadamard rotations are generated by applying a local magnetic field to the
boundary rung, while the Pauli-Z operation is generated by free evolution under
the finite-size splitting \(\Delta_0\). The corresponding gate time of the Hadamard operation is $t_\text{gate} = \frac{\sqrt{2}\pi}{2\Delta_0}$. At this time, we compute \(F_{\mathrm{gate}}\) and \(P_{\mathrm{leak}}\) by
comparing the full TEBD-evolved state with the ideal logical target state for several qubit sizes (number of rungs).
The resulting values are reported in Supplementary Table~\ref{tab:single_operations}.

\begin{table}
\centering
\setlength{\tabcolsep}{10pt}
\begin{tabular}{c|ccc|cc}
\hline
Qubit size & $B/J$ & $\Delta_{0}/J$ & $t_\text{gate} \cdot J$ 
& $F_{\mathrm{gate}}$ & $P_{\mathrm{leak}}$ \\
\hline
$L = 12$ & $8.53\cdot 10^{-2}$ & $9.71\cdot 10^{-2}$ & $22.8$ 
& $0.99656$ & $3.38\cdot 10^{-3}$ \\

$L = 16$ & $4.51\cdot 10^{-2}$ & $4.96\cdot 10^{-2}$ & $44.8$ 
& $0.99930$ & $6.99\cdot 10^{-4}$ \\

$L = 20$ & $2.38\cdot 10^{-2}$ & $2.58\cdot 10^{-2}$ & $86.2$ 
& $0.99950$ & $4.98\cdot 10^{-4}$ \\

$L = 24$ & $1.25\cdot 10^{-2}$ & $1.34\cdot 10^{-2}$ & $165.7$ 
& $0.99988$ & $1.24\cdot 10^{-4}$ \\

$L = 28$ & $6.52\cdot 10^{-3}$ & $6.96\cdot 10^{-3}$ & $319.2$ 
& $0.99997$ & $2.85\cdot 10^{-5}$ \\

$L = 30$ & $4.70\cdot 10^{-3}$ & $5.01\cdot 10^{-3}$ & $443.4$ 
& $0.99998$ & $1.57\cdot 10^{-5}$ \\
\hline
\end{tabular}
\caption{$|$ \textbf{Single-qubit Hadamard gate performance.} Gate performance for a single Haldane qubit as a function of the subsystem size
\(L\). The boundary magnetic field \(B\) is applied to the edge rung and chosen
to generate a Hadamard rotation in the logical subspace
\(\{|0\rangle,|1\rangle\}\). The finite-size splitting between the two logical
states is denoted by \(\Delta_0\), and the gate time is
\(t_{\mathrm{gate}}=t_H=\sqrt{2}\pi/(2\Delta_0)\). The gate fidelity
\(F_{\mathrm{gate}}\) and leakage probability \(P_{\mathrm{leak}}\) are obtained
from the full TEBD time evolution by projecting the evolved many-body state onto
the logical subspace. Times are reported in dimensionless units, with
\(\hbar=1\) and \(J=1\).}
\label{tab:single_operations}
\end{table}

\subsection*{Two-qubit entangling gate performance}
For the two-qubit protocol of Fig.~6, the two logical qubits are encoded in
neighboring Haldane subsystems separated by a rung-singlet boundary. The entangling operation is generated by an Ising-type interaction between the boundary rungs of the two subsystems,
\[
\hat H_{\mathrm{int}}
=
g_2 \hat S^{z(A)}_1 \hat S^{z(B)}_L .
\]
In the projected logical subspace this gives an effective
\(\sigma_x\otimes\sigma_x\) interaction, together with the finite-size
splittings of the two subsystems. Starting from the initial state \(|00\rangle\),
we evaluate the fidelity with the target Bell state $\frac{1}{\sqrt{2}}(\ket{00} + \ket{11})$, i.e the time where the
entanglement entropy between the two subsystems reaches its first maximum. The corresponding gate fidelity and leakage probability are computed for different system sizes and are reported in Supplementary Table~\ref{tab:double_operations}.
\\
For these protocols of realization of the Hadamard and entangling gate, the fidelity increases with system size mainly because the external parameters: $B$ and $g_2$, get smaller as the number of rungs increases. This is a consequence of the decrease in the gap between $\ket{0}$ and $\ket{1}$ as the qubit size increases.

\begin{table}[h!]
\centering
\setlength{\tabcolsep}{10pt}
\begin{tabular}{c|cc|cc}
\hline
Qubit size & $g_{2}/J$ & $t_{gate} \cdot J$ 
& $F_{\mathrm{gate}}$ & $P_{\mathrm{leak}}$ \\
\hline
$L_{A} = L_{B} = 12$ & $3.00\cdot 10^{-1}$  & $11.4$ & $0.97873$ & $2.06\cdot 10^{-2}$ \\

$L_{A} = L_{B} = 16$ & $1.64\cdot 10^{-1}$ & $22.4$ & $0.99074$ & $9.19\cdot 10^{-3}$ \\

$L_{A} = L_{B} = 20$ & $8.83\cdot 10^{-2}$ & $43.1$ & $0.99618$ & $3.81\cdot 10^{-3}$ \\

$L_{A} = L_{B} = 24$ & $4.67\cdot 10^{-2}$ & $82.9$ & $0.99894$ & $1.05\cdot 10^{-3}$ \\

$L_{A} = L_{B} = 28$ & $2.44\cdot 10^{-2}$ & $159.6$ & $0.99967$ & $3.24\cdot 10^{-4}$ \\

$L_{A} = L_{B} = 30$ & $1.76\cdot 10^{-2}$ & $221.7$ & $0.99986$ & $1.38\cdot 10^{-4}$ \\
\hline
\end{tabular}
\caption{$|$ \textbf{Two-qubit entangling gate performance.}
Gate performance for two neighboring Haldane qubits with equal subsystem sizes
\(L_A=L_B\). The qubits are separated by a rung-singlet boundary and coupled
through an edge-edge Ising interaction of strength \(g_2\). Starting from
\(|00\rangle\), the gate time \(t_{\mathrm{gate}}\) corresponds to the first
maximum of the entanglement entropy between the two subsystems, where the target
Bell state \((|00\rangle+|11\rangle)/\sqrt{2}\) is generated. The gate fidelity
\(F_{\mathrm{gate}}\) and leakage probability \(P_{\mathrm{leak}}\) are computed
from the full TEBD time evolution by projecting onto the two-qubit logical
subspace
\(\{|00\rangle,|01\rangle,|10\rangle,|11\rangle\}\). Times are reported in
dimensionless units, with \(\hbar=1\) and \(J=1\).}
\label{tab:double_operations}
\end{table}

\end{document}